\documentclass[aps,pre,reprint,superscriptaddress,twocolumn,showkeys,amsmath,amssymb,longbibliography]{revtex4-1}
\usepackage[english]{babel}
\usepackage{amsmath,amssymb,bbm,mathrsfs,bm,braket,color,graphicx,comment,amsfonts,dsfont}
\usepackage[colorlinks,citecolor=blue,urlcolor=blue]{hyperref}
\usepackage[normalem]{ulem}
\usepackage{comment}

\newcommand{\ave}[1]{\left\langle #1 \right\rangle}

\definecolor{purple}{rgb}{0.7, 0., 0.8}

\begin{document}

\title{The role of dimensionality and geometry in quench-induced nonequilibrium forces}

\author{M. R. Nejad}\email{mehrana.raeisiannejad@physics.ox.ac.uk}
\affiliation{The Rudolf Peierls Centre for Theoretical Physics, 1 Keble Road, Oxford, OX1 3NP, UK}

\author{H. Khalilian}
\affiliation{
School of Nano Science, Institute for Research in Fundamental Sciences (IPM) - P. O. Box 19395-5531,
Tehran, Iran
}

\author{C. M. Rohwer}
\affiliation{Department of Mathematics \& Applied Mathematics, University of Cape Town, 7701 Rondebosch, Cape Town, South Africa}
\affiliation{Max-Planck-Institut f\"ur Intelligente Systeme, Heisenbergstr.~3, 70569 Stuttgart, Germany}
\affiliation{IV.~Institut f\"ur Theoretische Physik, Universit\"{a}t Stuttgart,Pfaffenwaldring 57, D-70569 Stuttgart, Germany}

\author{A. G. Moghaddam}
\affiliation{Department of Physics, Institute for Advanced Studies in Basic Sciences (IASBS), Zanjan 45137-66731, Iran}
\affiliation{Research Center for Basic Sciences \& Modern Technologies (RBST), Institute for Advanced Studies in Basic Science (IASBS), Zanjan 45137-66731, Iran}


\begin{abstract}
{We present an analytical formalism, supported by numerical simulations, for studying forces that act on curved walls following temperature quenches of the surrounding ideal Brownian fluid. We show that, for curved surfaces, the post-quench forces initially evolve rapidly to an extremal value, whereafter they approach their steady state value algebraically in time. In contrast to the previously-studied case of flat boundaries (lines or planes), the algebraic decay for the curved geometries depends on the dimension of the system. Specifically, the steady-state values of the force are approached in time as $t^{-d/2}$ in d-dimensional spherical (curved) geometries. For systems consisting of concentric circles or spheres, the exponent does not change for the force on the outer circle or sphere. However, the force exerted on the inner circle or sphere experiences an overshoot and, as a result, does not evolve towards the steady state in a simple algebraic manner. The extremal value of the force also depends on the dimension of the system, and originates from the curved boundaries and the fact that particles inside a sphere or circle are locally more confined, and diffuse less freely than particles outside the circle or sphere.
}
\end{abstract}

\maketitle

\section{Introduction}
\label{sec:intro}

Objects immersed in fluctuating media can experience forces for a variety of reasons. The prototypical example is that of fluctuation-induced forces (FIFs), also referred to as Casimir and van der Waals forces
\cite{casimir,fisher1978wall,kardar99,hertlein2008direct}, which can arise because the objects modify fluctuation modes of the medium in the presence of long-ranged correlations. Interestingly, these FIFs exhibit many universal properties that are independent of details of the medium and the nature of the fluctuations (e.g., quantum or thermal), and emerge in different contexts ranging from atomic and molecular physics condensed matter, and biology to material science, chemistry and biology \cite{bordag2009advances,podgornik16b,long-range,parsegian2006handbook}. In classical systems at thermal equilibrium, FIFs generally stem from long-ranged correlations in the vicinity of critical points. In nonequilibrium situations, however, long-ranged correlations can emerge much more generally \cite{spohn90,sachdev90}. In particular, the conservation of different global quantities (e.g., the number of particles) can lead to constrained nonequilibrium dynamics which subsequently give rise to long range correlations and FIFs \cite{spohn1983,dorfmankirkpatricksengers1994,mukamelkafri1998}. For classical systems, nonequilibrium FIFs have been previously explored in the steady-states of externally driven systems, e.g., in the presence of temperature or density gradients \cite{kirkpatrick-prl,kirkpatrick-pre,aminovkardarkafri2015,soto2003}. Recently, the role of sudden changes (quenches) of the temperature or of other system parameters in inducing a transient nonequilibrium states and FIFs has been studied by various groups ~\cite{gambassi2008EPJB,deangopinathan2010PRE,rohwer17,rohwer18,gross2018surface,rohwer19,gross2019ModelBcrit}.
 
While nonequilibrium dynamics following quenches has attracted much attention in the context of interacting quantum systems \cite{calabrese},  various interesting phenomena also emerge in classical quench dynamics. In particular, it has been found that aside from FIFs, post-quench dynamics of the \textit{density in classical fluids} can give rise to additional forces on immersed objects or surfaces \cite{rohwer17,rohwer18,Khalilian2020}. Such density-induced forces (DIFs) exist (and become longer-ranged) even in non-interacting (ideal) fluids. In contrast, the fluctuation forces discussed above rely on correlations, and disappear in the absence of interactions between fluid particles.

It is well-established that geometry and dimensionality play an important role in equilibrium Casimir-type (FIF) systems \cite{kardar-geometry}. However, for forces induced by sudden quenches of temperature (or of activity in active fluids), these effects have not been considered thus far. In this paper, we study the transient forces following a quench of temperature in an ideal (Brownian) fluid confined by curved surfaces or inclusions.
In Ref.~\cite{rohwer18} such post-quench DIFs were considered for flat walls immersed in an ideal fluid, and were shown to be independent of the dimension of the system and the boundaries:  the force on the walls approaches its equilibrium value algebraically in time as ${t}^{-1/2}$. Here we aim to understand how this picture is modified by the curvature of the boundaries. We approach the problem with a theory for diffusion in curved geometries, as well as through explicit simulations of Brownian dynamics in the given setup.
While our study applies to fluids confined by spherical or curved surfaces, it may also be of interest for the dynamics and behavior of fluid droplets in a quenched medium.

The paper is organized as follows: We first introduce the model system and simulation details in Sec.~\ref{sec:model}. Using a coarse-grained description, we determine the analytical solutions for the post-quench dynamics of the density field of Brownian particles,
from which the forces on curved walls can be also obtained. Then, in Secs.~\ref{sec:d=2} and \ref{sec:d=3}, numerical and analytical results are presented for the pressure and the force exerted on circular and spherical boundaries, respectively. 
Finally, in Sec.~\ref{sec:conc}, we close the paper with a summarising discussion and conclusion.

\section{Model and simulation details}\label{sec:model}
\begin{figure}[t!]
\centering
\includegraphics[width=0.7\linewidth]{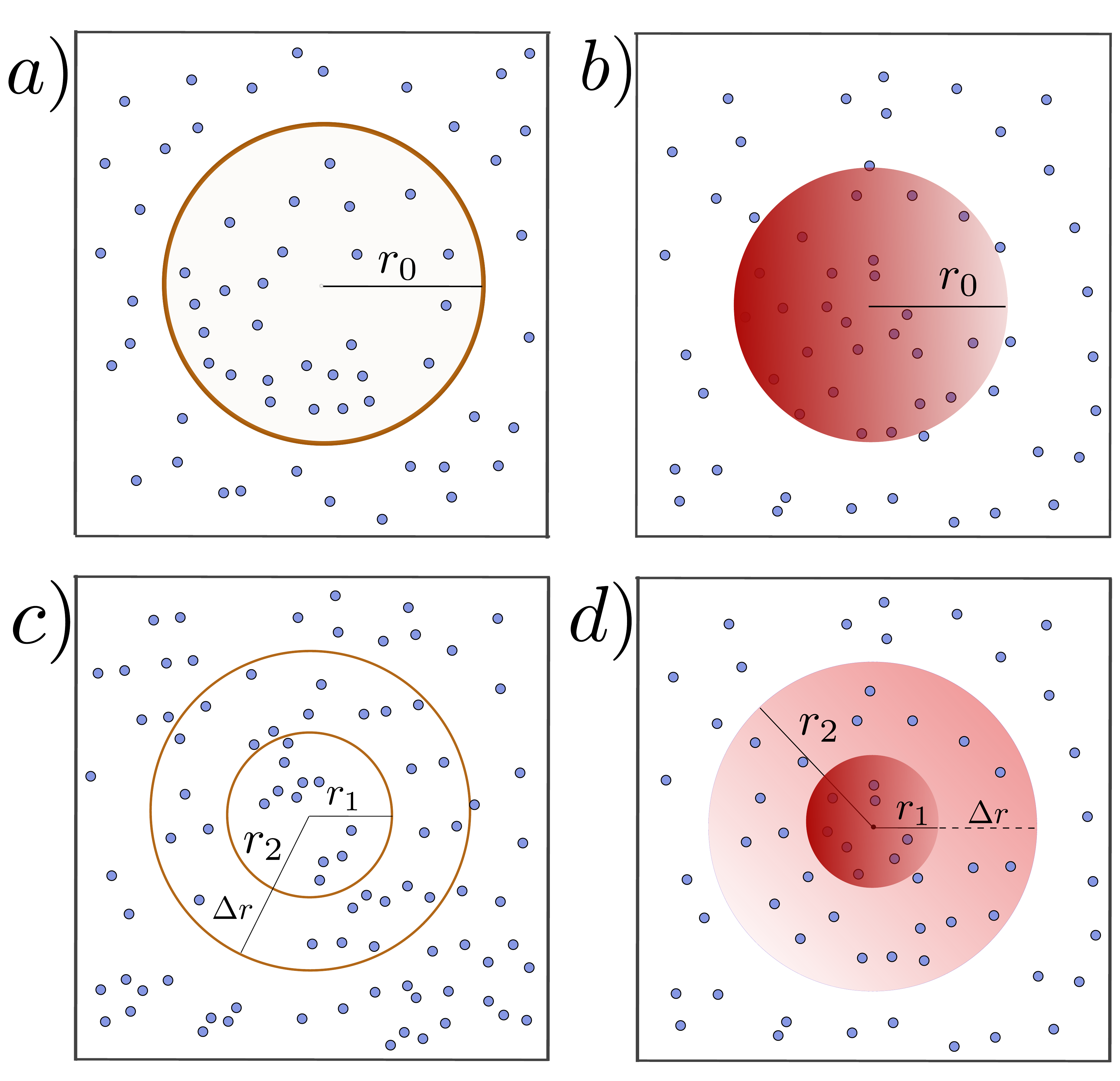}
\caption{
Panels (a) and (c) show a system with spherical symmetry in 2D; (b) and (d) show the same in 3D. The medium (black dots) is an ideal gas of passive Brownian particles. Panels (a) and (b) correspond to a circle and a sphere with radius $r_0$, respectively. Panels (c) and (d) show a Brownian gas inside and outside of two inclusions with radii $r_1$ and $r_2$, for circular and spherical cases, respectively.
}
\label{fig1}
\end{figure}

We study dynamics of a non-interacting Brownian fluid, inside and outside of a sphere in $d$ dimensions, after a quench in temperature. In particular, we consider the pressure and force exerted on the confining spherical walls in terms of time dependence and their scaling behavior. By symmetry of the sphere, there is no preferred direction for currents at the center of the sphere. In simulations, we model walls of the sphere by a repulsive quadratic potential. Dynamics of the system is described by over-damped Langevin equations:
\begin{eqnarray}\label{onee}
&&\frac{d \textbf{r}_i}{d t}= -\mu \frac{d V_{\rm wall}}{d r}|_{\textbf{r}_i} \hat{r}+\boldsymbol{\eta}_i(t),
\end{eqnarray}
Here, $\textbf{r}_i$ is the position of $i$th particle ($i=1,2,\ldots, N$), $\mu$ the mobility of the particles, and $V_{\rm wall}$ represents the confining wall potential,
\begin{equation}\label{twoo}
V_{\rm wall}(r)=\frac{\lambda}{2} [\Theta(r-R_d)(r-R_d)^2],
\end{equation}
where $R_d$ is the radius of the $d$-dimensional confining spherical boundaries. From now on and for the sake of clarity, we denote the radius in the case of only one sphere (circle) with $r_0$, whereas in the cases of two  spheres (circles), the two radii are denoted by $r_1$ and $r_2$ with a difference $\Delta r$ (See Fig. \ref{fig1}).  

In Eq.~\eqref{twoo}, $\Theta(x)$ represents Heaviside step function, which is equal to one for $x>0$ and vanishes otherwise. The strength of the quadratic potential is given by $\lambda$. In Eq.~\eqref{onee}, $\boldsymbol{\eta}_i$ is Gaussian white noise obeying 
\begin{equation}
\ave{\eta_{i\alpha}(t) \eta_{j\beta}(t^{\prime})}= D \delta(t-t^{\prime}) \: \delta_{ij}\delta_{\alpha \beta},
\end{equation}
where $D$ is the diffusion coefficient of the particles, and $\eta_{i\alpha}$ is  the $\alpha$th component of the noise exerted on particle $i$.

Using Eqs.~\eqref{onee} and \eqref{twoo}, we simulated the dynamics of the system of passive Brownian particles inside a spherical geometry in two or three dimensions, following the temperature quench. In all simulations, we average over $N=10^5$ particles to obtain the pressure on the walls.
A forward Euler method was employed to integrate Eq.~\eqref{onee} in order to study the post-quench dynamics of each particle. In simulations, we instantaneously change the temperature by changing the diffusion coefficient of the particles in Eq. \eqref{onee}. The force exerted on the wall is then the sum of all single-particle forces as
\begin{equation}\label{three}
{\bf F}= \sum_{i=1}^{N}  \frac{d V_{\rm wall}}{d r}|_{\textbf{r}_i} \hat{\bf r}.
\end{equation}

We also provide a coarse-grained theoretical description based on the Smoluchowski equation, and study the evolution of the density of the non-interacting Brownian particles following the quench. The particle density is defined as $\rho(\textbf{r},t)= \sum_{i=1}^N \delta(\textbf{r}-\textbf{r}_i(t))$. The Langevin equation \eqref{onee} leads to a spherically symmetric coarse-grained dynamics given by
\begin{equation}\label{five}
\partial_{t} \rho(r,t) = \frac{D}{r^{d-1}}\partial_{r}\big[r^{d-1} \partial_r \rho(r,t)\big] ,
\end{equation}
in $d$ dimensions.
In this coarse-grained picture, walls are assumed to impose no-flux boundary conditions at $r=R_d$:
  \begin{equation}\label{ix}
\partial_{r} \rho(r,t)|_{r=R_d}=0.
\end{equation}

Before the quench, the system is in a steady state described by the canonical Boltzmann weight $\rho_I \propto \exp[- V_{\rm wall}(r)/(k_B T_I)]$, where $k_B$ is Boltzmann's constant. At time $t=0$, we change the temperature to $T=T_F$ instantaneously. After this quench, the system evolves to a new equilibrium state which is described by  $\rho_F \propto \exp[- V_{\rm wall}(r)/(k_B T_F)]$. As discussed in Ref.~\cite{rohwer18}, the quench modifies the boundary layer of particles close to the walls, because the penetration depth of the particles into the wall potential is temperature dependent. We can therefore write the post-quench particle density as the sum of a homogeneous contribution, representing the region outside of the wall potential, and a time- and position-dependent excess density, 
  \begin{equation}\label{sixx}
\rho(r,t) = \rho_0 + \Delta \rho(r,t).
\end{equation}
In the coarse-grained description, we consider strongly repulsive walls, so that the initial excess density after the quench is approximated as a $\delta$-function at the boundary,
 \begin{equation}\label{six}
\Delta \rho(r,t=0)= \alpha_{d} \rho_0 \delta(r-R_d).
\end{equation}
Here $\rho_0$ and $\alpha_d$ scale with length as $[\rho_0] = 1/\ell^d$ and $[\alpha_d] = \ell$, respectively. Indeed, $\alpha_d$ corresponds to the change of the width of the boundary layer induced by the quench, and can be computed by integrating $\rho_I$ and $\rho_F$ over the relevant volume and enforcing conservation of the particle number. We calculate this parameter analytically in Appendix \ref{ap1}.
After the quench, this adsorbed (or desorbed, depending on whether the quench is to a higher or lower temperature) layer of particles at the boundary diffuses into the system. This gives rise to dynamics of post-quench forces and pressures acting on the boundaries.
It turns out that $\alpha_d$ is independent of the system's dimension; we therefore use the notation $\alpha$ throughout instead. Using separation of variables, we can solve Eq.~\eqref{five} and find the evolution of the excess density in time. The ideal gas law then provides the corresponding instantaneous pressure exerted by Brownian particles on the $d$-dimensional spherical wall after the quench,
 \begin{equation}\label{seven}
P(r=R_d,t)=k_B T_F [\rho_0+\Delta \rho(r=R_d,t)].
\end{equation}
Solving Eqs.~\eqref{five}, \eqref{ix} and \eqref{six} for the region \textit{outside} of the sphere, we can analogously find the evolution of the density of Brownian particles on the exterior of the sphere after the quench (similar no-flux boundary conditions apply at the wall). Subtracting the outside from the inside pressure exerted on the sphere, one obtains the force exerted per area $A_d$ on the sphere following the quench:
 \begin{equation}\label{eight}
\frac{F(t)}{A_d}=k_B T_F [\Delta \rho_{\rm in}(R_d,t)-\Delta \rho_{\rm out}(R_d,t)].
\end{equation}
In Eq.~\eqref{eight}, $ \Delta \rho_{\rm in/out}$ represents the excess density inside/outside the sphere. For time, pressure, and force, we define the dimensionless variables 
 \begin{equation}\label{e8}
\bar{t}= \frac{D t}{R_d^2}, 
\:\:\: 
\bar{P}(t)=\frac{P(t) R_{d}}{\alpha_{d} \rho_0 k_B T_F}, \:\:\:
\bar{F}(t)=\frac{F(t) R_{d} }{\alpha_{d} \rho_0 k_B T_F A_d}.
\end{equation}
Here $R_d$ takes the values $r_0$, $r_1$, or $r_2$, depending on the considered geometry and the wall for which we calculate the pressure and force.

\section{Quench in circular geometries ($d=2$)}\label{sec:d=2}

In this section we discuss the post-quench pressure and force  acting on a single circular wall or on two concentric circles of different radii. 

\subsection{Inside the circle}
Setting $d=2$ in Eq.~(\ref{five}), we have
\begin{equation}\label{30}
\partial_t \rho(r,t)=\frac{D}{r}\partial_r [r \partial_r \rho(r,t)].
\end{equation}
We then use separation of variables to write density as $\Delta \rho(r,t)=X(r) T(t)$. Putting this density back to Eq. (\ref{30}) we have
\begin{equation}\label{31}
\frac{\partial_t T(t)}{T}= \frac{D}{X r}(\partial_r X+r \partial_r^2 X)=-\zeta^2.
\end{equation}
Solving for $X$ and $T$ we find
\begin{equation}\label{32}
T(t)= e^{-\zeta^2 t},\:\:\:\:\:X(r)=c_1 J_0(\frac{\zeta r}{\sqrt{D}}) +c_2 Y_0(\frac{\zeta r}{\sqrt{D}}),
\end{equation}
where $J_n$ and $Y_n$ are $n$th order Bessel functions of the first and second kind, respectively~\cite{abramowitz}. The solution found in Eq.~(\ref{32}) should be finite inside the circle, so $c_2=0$. The no-flux boundary condition at the wall translates into
\begin{equation}\label{33}
\partial_r X(r)|_{r=r_0}=0,
\end{equation}
which implies that the parameter $\zeta$ takes on discrete values. Defining $\beta=\zeta r_0/\sqrt{D}$, the allowed values for $\beta$ are the positive roots of the Bessel function,
\begin{equation}\label{34}
J_1(\beta_n) =0,  ~~~ n=1,2,3,\dots~.  
\end{equation}
We can then write the time-dependent evolution of the density in the form of an infinite series,
\begin{equation}\label{35}
\Delta \rho_{\rm in}(r,t)= \sum_{n=1}^{\infty} c_n e^{-\beta_n^2 \bar{t}} J_0(\beta_n \frac{r}{r_0})+c_0,
\end{equation}
using dimensionless time $\bar t$ as defined in Eq.~\eqref{e8} with $R_d=r_0$.
The appropriate initial condition for this problem (which involves the excess density at the wall) takes the form $\Delta\rho_{\rm in}(r,t=0)=\alpha \rho_0 \delta(r-r_0)$, and
the constant $c_0$ can be found by integrating over the disk and enforcing particle conservation:
\begin{equation}\label{36}
\int_{0}^{r_0} r \: dr \: \big[\sum_{n=1}^{\infty} c_n J_0(\beta_n \frac{r}{r_0})+c_0\big] = \int_{0}^{r_0} r \: dr \: \alpha \rho_0 \delta(r-r_0) ,
\end{equation}
which gives $c_0=2 \rho_0 \alpha/r_0$. To find the remaining coefficients $c_n$, we use the following orthogonality relation for the Bessel function~\cite{abramowitz}:
\begin{eqnarray}\label{37}
\int_{0}^{1} x dx J_0(\beta_n x)J_0(\beta_m x)=\delta_{nm}\frac{[J_0(\beta_n)]^2}{2},
\end{eqnarray}
with $\beta_n$'s given by Eq.~(\ref{34}). This yields
\begin{equation}\label{38}
c_m=\frac{\alpha \rho_0 \: r_0 J_0(\beta_m)}{\int_0^{r_0} r\, dr \, [J_0(\beta_m \bar{r})]^2 }=\frac{2 \alpha \rho_0 }{ r_0 J_0(\beta_m) }
\end{equation}
Putting everything together we find the density on the circle as
\begin{equation}\label{39}
\Delta \rho_{\rm in}(r_0,t)= \bigg(1+\sum_{n=1}^{\infty} e^{- \beta_n \bar{t}}  \bigg)\frac{2 \alpha \rho_0}{r_0}.
\end{equation}

\subsection{Outside the circle}
For the evolution of density outside the circle we similarly use Eqs.~(\ref{32}) and (\ref{33}), from which we obtain the following result:
\begin{eqnarray}\label{40}
&&\Delta \rho_{\rm out}(r,t)= \int_{0}^{\infty} d\gamma\, c(\gamma)  \:{\cal G}(\gamma,\frac{r}{r_0}) \:e^{-\gamma^{2}\bar{t}}\,,  \nonumber \\&&
{\cal G}(\gamma,x)=Y_1( \gamma) J_0(\gamma \:x)-J_1(\gamma) Y_0(\gamma \:x)\,. 
\end{eqnarray}
To find the coefficients $c(\gamma)$, we insert the above result in the initial condition written in the form 
\begin{eqnarray}
&&\int_{r_0}^\infty r \: dr \: \Delta \rho_{\rm out}(r,t=0) {\cal G}(\gamma^{\prime},\frac{r}{r_0}) \nonumber\\
&&~~~~ =\alpha \rho_0 \int_{r_0}^\infty r \: dr \: \delta(r-r_0)  {\cal G}(\gamma^{\prime},\frac{r}{r_0}) , 
\end{eqnarray}
and evaluate the integrals by invoking the orthogonality condition 
\begin{equation}
\int_1^\infty x dx \: {\cal G}(\gamma,x){\cal G}(\gamma^{\prime},x) = \:\delta(\gamma-\gamma^{\prime}) \frac{[J_1( \gamma) ]^2+[Y_1( \gamma) ]^2}{\gamma}.
\end{equation}
This finally leads to 
\begin{equation}\label{41}
c(\gamma)= \frac{-2 \alpha \rho_0}{ \pi r_0 \big[ J_1^2(\gamma)+Y_1^2(\gamma)\big]}.
\end{equation}
Then, the density on the boundary of the circle reads
\begin{eqnarray}\label{40h}
\Delta \rho_{\rm out}(r_0,t)= \int_{0}^{\infty} d\gamma\, \frac{4 \alpha \rho_0 \: e^{-\gamma^{2}\bar{t}}}{\gamma \pi^2 r_0 \big[ J_1^2(\gamma)+Y_1^2(\gamma)\big]}.  
\end{eqnarray}
In the long time limit, this excess density decays as
\begin{eqnarray}\label{40j}
\Delta \rho_{\rm out}(r_0,t\to\infty)\approx \frac{\alpha \rho_0}{2 r_0} \: \bar{t}^{-1}.
\end{eqnarray}
At long times after the quench, the excess pressure exerted on the circle from outside therefore approaches zero as $\bar{t}^{-1}$.

\begin{figure}[t!]
\includegraphics[width=0.7\linewidth]{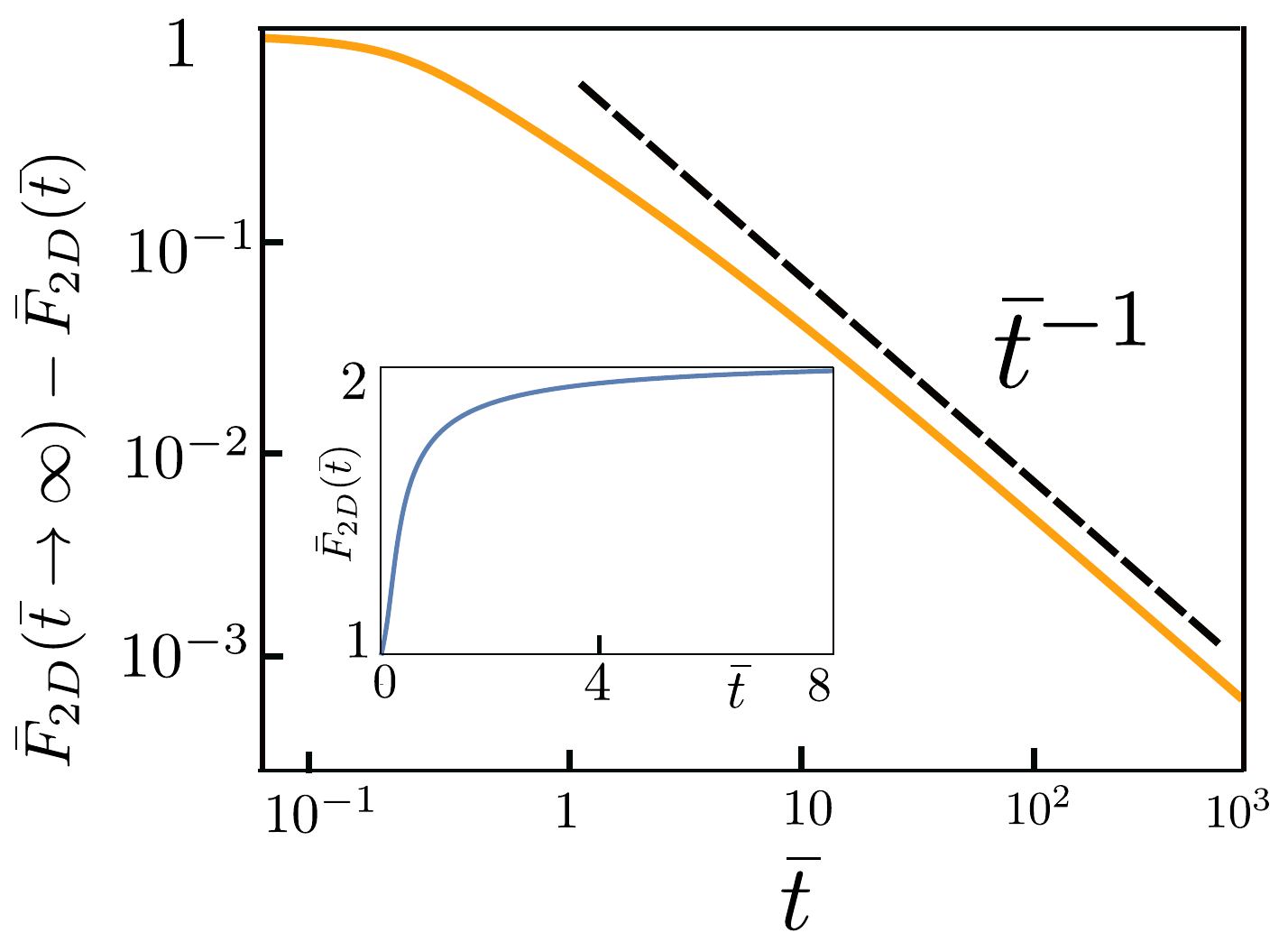}
\centering
\caption{Analytical results for the temporal approach of the force acting on the circle towards its new steady-state value, shown on logarithmic (main panel) and linear (inset) scales.
}
\label{fignew}
\end{figure}

Having computed the density of the particles inside and outside the circle, we can use Eq.~\eqref{eight} to find the force exerted on the circle after the quench. Figure \ref{fignew}, shows that the force approaches its new steady-state as $\bar{t}^{-1}$ in time. 

As can be seen in the inset of the figure, the force exerted on the circle jumps very rapidly to the value $\bar{F}_{2D}=1$ short after the quench. This jump is a consequence of the curved geometry of the circle and a clear manifestation of the nonequilibrium character of the quench-induced dynamics. The curvature of the wall implies that diffusion of the excess particles inside the circle occurs in a confined environment, in contrast to the outside region. As a result, and assuming initial excess densities which are entirely localized at the boundaries, this finite force is reached immediately (and discontinuously in time) after the quench. 
However, this observation is a consequence of the coarse-graining assumptions: if we attribute a finite small width $\epsilon$ to the initial excess density (rather than the delta function), the force becomes finite after a very short time $\bar{t}\gtrsim(\epsilon/r_0)^2$ (see Appendix \ref{ap2} for more details). 

\subsection{Quench effect on the medium between two circles}

We now consider a system of non-interacting Brownian particles confined between two circles with radii $r_1$ and $r_2$, as shown in Fig.~\ref{fig1} (c). Using the solution found for the density in Eq.~\eqref{32}, and imposing no-flux boundary conditions on both circles, $\partial_r X(r)|_{r=r_1,r_2}=0$, we can find discrete values for the parameter $\beta^{\prime}$ from the following equation:
\begin{equation}\label{41B}
\frac{J_1(\beta^{\prime}_n r_1)}{Y_1(\beta^{\prime}_n r_1)}=\frac{J_1(\beta^{\prime}_n r_2)}{Y_1(\beta^{\prime}_n r_2)} ,~~~ n=1,2,3,\dots~. 
\end{equation}

The time-dependent solution for the density can be written as
\begin{eqnarray}\label{41c}
&& \Delta\rho(r,t)= \sum_{n=1}^{\infty} c_n f_n(r) e^{-\beta_n^{\prime^2}D t} + c_0,\\&&
f_n(r)=J_0(\beta^{\prime}_n r) Y_1(\beta^{\prime}_n r_2)-J_1(\beta^{\prime}_n r_2) Y_0(\beta^{\prime}_n r),\nonumber
\end{eqnarray}
where the summation runs over all positive solutions of $\beta^{\prime}$ in Eq.~\eqref{41B}. 
The initial condition, which corresponds to excess particle layers at both surfaces, has the form of $\Delta\rho(r,t=0) = \alpha \rho_0 \big[\delta(r-r_1)+ \delta(r-r_2)\big]$, and the constant $c$ can be found by calculating the following integrals related to conservation of the particle number:
\begin{eqnarray}\label{41d}
&&\int_{r_1}^{r_2} r \: dr \: [c_n f_n(r) + c_0] \nonumber\\&&= \alpha \int_{r_1}^{r_2} r \: dr \: [ \delta(r-r_1)+ \delta(r-r_2)].
\end{eqnarray}
This gives
\begin{equation}\label{41e}
c_0=\frac{2\alpha \rho_0}{r_2-r_1}. 
\end{equation}
For calculating coefficients $c_n$ we need to calculate below integrals:
\begin{eqnarray}\label{41f}
&& \int_{r_1}^{r_2}  r \: dr  \: \alpha \rho_0 \big[ \delta(r-r_1)+ \delta(r-r_2)\big] f_m(r) \nonumber \\&&=\int_{r_1}^{r_2} r \: dr \: \big[ \sum_{n=1}^{\infty} c_n f_n(r) + c_0 \big] f_m(r) ,
\end{eqnarray}
which gives
\begin{equation}\label{41g}
c_n=\frac{2 \alpha \rho_0 }{r_2 f_n(r_2)-r_1 f_n(r_1)}.
\end{equation}
To calculate the coefficients $c_n$, we have used the orthogonality relation
\begin{eqnarray}\label{orth2c}
\int_{r_1}^{r_2} r \: dr \: f_m(r) f_n(r)=\frac{\delta_{m n}}{2}
 [r_2^2 f_n^2(r_2)-r_1^2 f_n^2(r_1)].
\end{eqnarray}

\begin{figure}[t!]
\centering
\includegraphics[width=0.99\linewidth]{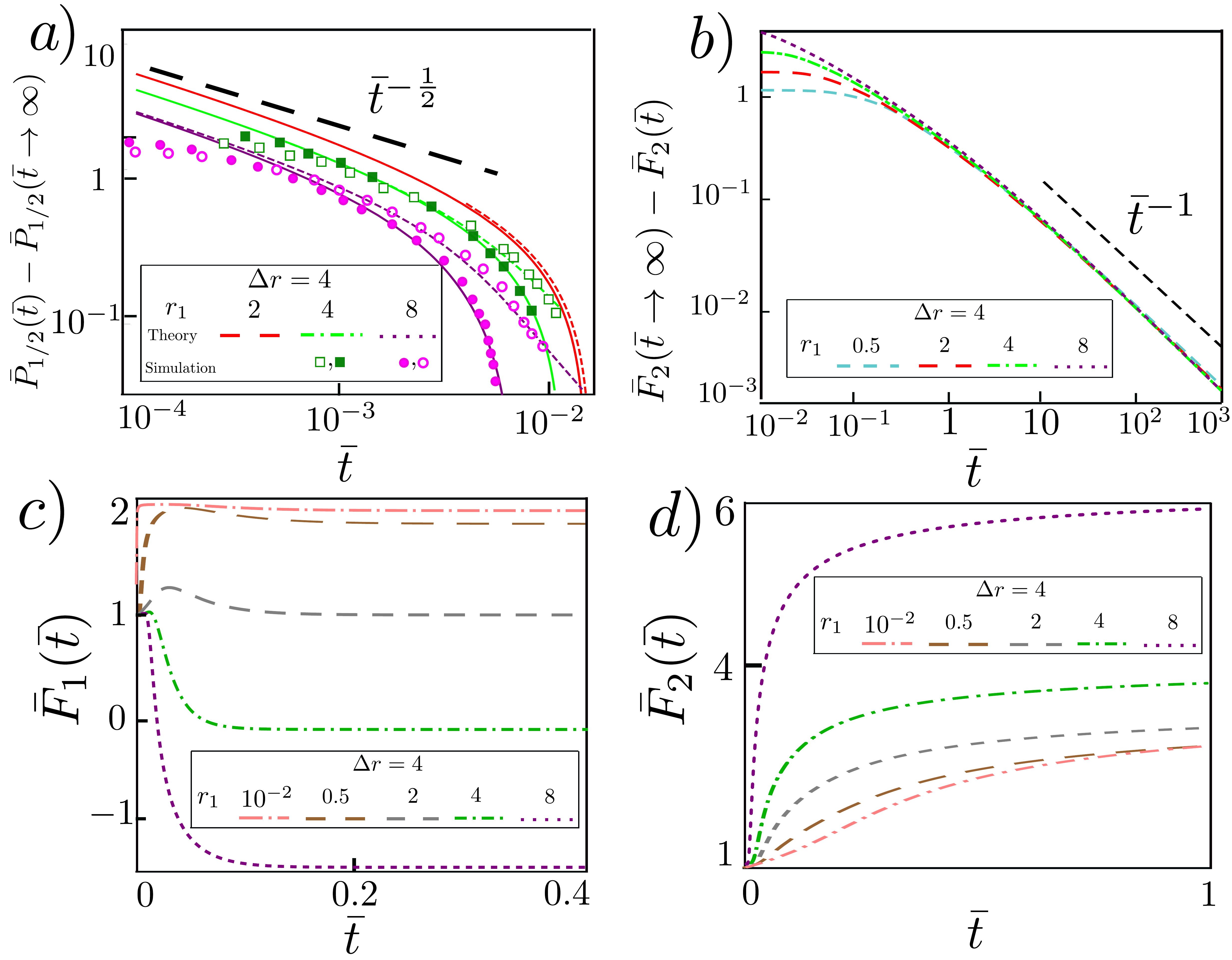}
\caption{
(a) Time evolution of the pressure exerted on the smaller ($\bar P_1$) and larger ($\bar P_2$) circle, measured with respect to their steady-state value. The solid lines and filled symbols (dashed line and hollow symbols) correspond to $\bar P_1$ ($\bar P_2$).
(b) Evolution of the force exerted on the larger circle towards its steady-state value. Panels (c) and (d) show forces exerted on the small and large circles, respectively. Forces are obtained from the analytical calculations.
}
\label{fig2cir}
\end{figure}

The post-quench dynamics of pressure and forces for two concentric circles are represented in Fig.~\ref{fig2cir}.
Both analytical and simulation results, shown in Fig.~\ref{fig2cir}(a), indicate that the pressures $\bar{P}_1$ and $\bar{P}_2$ exerted on the small and large circles by the confined particles, scale as $\bar{t}^{-{1}/{2}}$ at short times after the quench. This behavior stems from the fact that at very short times the excess particle density is still very localized near the walls, and its evolution is effectively described by a one-dimensional diffusion along the direction locally normal to the wall. As will become clear in the next section, the $\bar{t}^{-{1}/{2}}$ behavior of the pressure also occurs in a spherical geometry and can be considered as a universal short-term characteristic of the pressure in all dimensions and geometries. 

We now turn to the force exerted on each circle. As shown in Fig.~\ref{fig2cir}(b), the force exerted on the large circle approaches its steady-state as $\bar{t}^{-1}$ at long times.
This behavior is the same as for the force on a single circle (Fig. \ref{fignew}), because the late-time behavior is dominated by the density relaxation in the infinite outside medium.
The early time behavior of the force on the smaller circle is also similar to the case of a single circle, and, as Fig. \ref{fig2cir}(c) shows, the force jumps to ${\bar F}_1=1$ immediately after the quench. 
However, the late time behavior differs significantly: after 
an initial overshoot, which becomes stronger and takes place at earlier times for larger $r_1$, a steady-state behavior is approached. 
To understand the non-monotonic behaviour of the force exerted on the smaller circle, we note that due to the curvature of the walls, at small times after the quench, the particles from inside the circle move towards the walls more than those particles between the circles. Correspondingly, the force exerted on the smaller circle is positive. At later times, the particles which were initially located on the exterior circle (at $r= r_2$) find time to diffuse and reach the smaller circle and, as a result, the force exerted on the smaller circle decreases. This explains the overshoot seen in Fig.~\ref{fig2cir}(c).
Upon varying $r_1$ while keeping $\Delta r$ fixed,
the steady-state force also changes and even undergoes a sign change for large $r_1$ depending on the balance between the particles inside / outside the small circle. In contrast to $\bar{F}_1$, the force on the outer circle is always positive (towards outside) and monotonically increases, as shown in Fig. \ref{fig2cir}(d).

\section{Quench in spherical geometries ($d=3$)}\label{sec:d=3}

In this section we investigate the phenomena considered in Sec.~\ref{sec:d=2}, but in three spatial dimensions.

\subsection{Inside the sphere}
\begin{figure}[t!]
\centering
\includegraphics[width=0.7\linewidth]{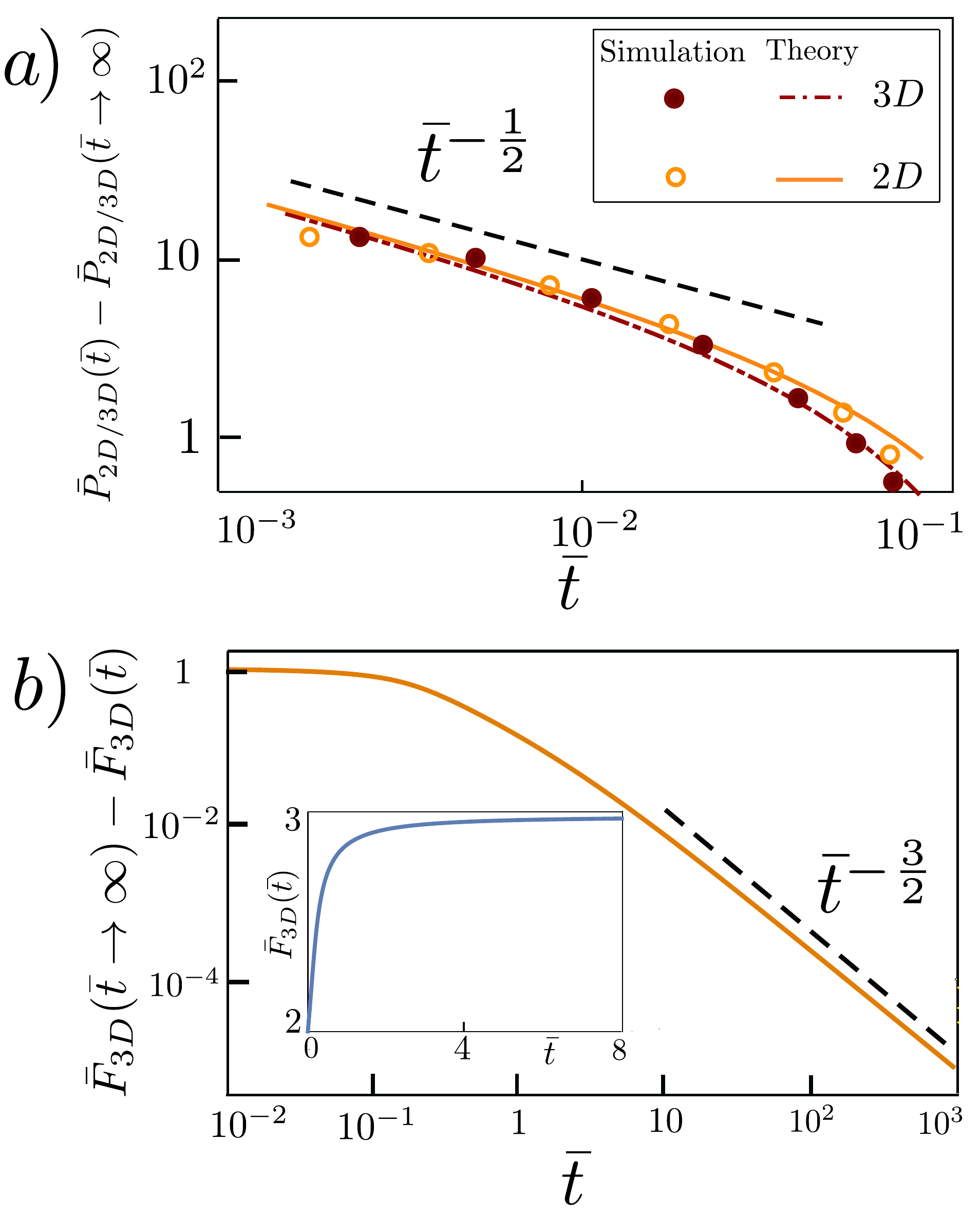}
\caption{(a) Short-term behavior of the pressure exerted on the circle (2D) and sphere (3D) from inside, towards its steady-state value. (b) Approach of the force exerted on the sphere towards its steady state value following the quench. 
}
\label{fig4}
\end{figure}

Dynamics of a diffusive system inside a sphere ($d=3$) can be written as
\begin{equation}\label{10}
\partial_t \rho(r,t)=\frac{D}{r^2}\partial_r\big[r^2 \partial_r \rho(r,t)\big].
\end{equation}
Again we use separation of variables for the excess density, $\Delta\rho(r,t)={\cal R}(r) T(t)$. Inserting this into Eq.~(\ref{10}) we find
\begin{equation}\label{11}
\frac{\partial_t T(t)}{T}= \frac{D}{{\cal R} r^2}\big(2 r \partial_r {\cal R}+r^2 \partial_r^2 {\cal R} \big)=-\zeta^2.
\end{equation}
Solving for $R$ and $T$ yields
\begin{equation}\label{12}
T(t)= e^{-\zeta^2 t},\quad{\cal R}(r)=\frac{A}{r} \cos\big(
\frac{\beta r}{r_0}\big)  +\frac{B}{r} \sin\big(
\frac{\beta r}{r_0}\big) ,
\end{equation}
where $\beta=\zeta r_0/\sqrt{D}$.
For the interior of a sphere, the first term in the solution for $R$ in Eq.~\eqref{12} diverges at $r=0$, so that one must put $A=0$. The no-flux boundary condition on the interior surface of the sphere gives
\begin{equation}\label{13}
\partial_{r} {\cal R}(r)|_{r=r_0}=0.
\end{equation}
Again we find discrete values for the parameter $\beta$, which are now the positive roots of
\begin{equation}\label{14}
\beta_n =\tan \beta_n, ~~~ n=1,2,3,\dots~. 
\end{equation}
The time-dependent solution for the excess density is then
\begin{equation}\label{15}
\Delta\rho_{\rm in}(r,t)= \sum_{n=1}^{\infty} c_n  \frac{\sin (\beta_n r/r_0)}{r}\,e^{- \beta_n^2 \bar{t}} +c_0.
\end{equation}
In Eq.~(\ref{15}), the summation is over all positive solutions of $\beta$ in Eq.~(\ref{14}). 
The initial condition $\Delta\rho_{\rm in}(r,t=0)=\alpha \rho_0 \delta(r-r_0)$ allows us to find the constant $c_0$ in analogy to the 2D case by calculating 
\begin{equation}\label{16}
\int_{0}^{r_0}  r^2 dr \:\Delta\rho_{\rm in}(r,t=0) = \int_{0}^{r_0} r^2 dr \: \alpha \rho_0 \delta(r-r_0),
\end{equation}
which gives $c_0=3 \alpha \rho_0/r_0$. The coefficients $c_n$ are obtained by applying the orthogonality condition
\begin{eqnarray}\label{16b}
\int^1_0 dx \, \sin(\beta_m x) \sin(\beta_n x) = \delta_{nm}\frac{\sin^2\beta_n}{2}
\end{eqnarray}
in the following relation:
\begin{eqnarray}\label{17}
&&\int_{0}^{r_0} r \: dr \: \Delta\rho_{\rm in}(r,t=0)  \sin(\beta_m r/r_0) \nonumber \\ && =\alpha \rho_0 \int_{0}^{r_0} r \: dr \: \delta(r-r_0) \sin(\beta_m r/r_0).
\end{eqnarray}
One finds
\begin{equation}\label{18}
c_m=\frac{2 \alpha \rho_0}{\sin \beta_m}.
\end{equation}
Finally the density on the interior surface of the sphere is obtained as
\begin{equation}\label{19}
\Delta\rho_{\rm in}(r=r_0,t)=\big(3+2\sum_{n=1}^{\infty} e^{- \beta_n^2 \bar{t}} \big)\frac{ \alpha \rho_0}{r_0}.
\end{equation}
The summation is over all positive solutions of $\beta$ in Eq. (\ref{14}).  
Using Eqs.~\eqref{39} and \eqref{19}, we can now calculate the pressure exerted from inside on the circle and the sphere, respectively. The scaling behavior is shown in Fig.~\ref{fig4}(a): at small times after the quench, the pressure decays towards its steady state value as $\bar{t}^{-1/2}$, similar to the case of circle ($d=2$).
As discussed in the previous section, this is a universal short-term characteristic of quench-induced pressure, and is independent of the dimensionality of the system.

\subsection{Outside the sphere}
Using Eqs.~\eqref{12} and \eqref{13}, the time-dependent excess density outside the sphere can be found:
\begin{eqnarray}\label{20}
&&\Delta\rho_{\rm out}(r,t)=\int_{0}^{\infty} d\gamma \, c(\gamma)  {\cal K}\big(\gamma^{\prime},\frac{r}{r_0}\big) e^{-\gamma^2 \bar{t}} ,\nonumber
\\ && {\cal K}(\gamma, x)= \frac{\gamma \cos [\gamma(x-1)]+\sin [\gamma(x-1)]}{x}.
\end{eqnarray}
The unknown coefficient $c(\gamma)$ is fixed by the initial condition along with Eq.~(\ref{20}). To this end we need to evaluate the integral relation 
\begin{eqnarray}\label{20t}
&&\int_{r_0}^{\infty} r^2 dr \: \Delta\rho_{\rm out}(r,t=0)\, {\cal K}\big(\gamma^{\prime},\frac{r}{r_0}\big)\nonumber
\\ &&~~~= \int_{r_0}^{\infty} r^2 dr \: \alpha \rho_0 \delta(r-r_0) \, {\cal K}\big(\gamma^{\prime},\frac{r}{r_0}\big).
\end{eqnarray}
Using the orthogonality relation
\begin{eqnarray}\label{20ortho}
\int_{1}^{\infty} dx x^2 {\cal K}(\gamma,x){\cal K}(\gamma^{\prime},x)=\frac{\pi}{2}\delta(\gamma-\gamma^{\prime})(1+\gamma^2),
\end{eqnarray}
we find
\begin{equation}\label{21}
c(\gamma)= \frac{2 \alpha \rho_0}{\pi r_0} \frac{\gamma}{1+\gamma^2}.
\end{equation}
The density on the exterior of the sphere takes the form
\begin{eqnarray}\label{22}
\Delta\rho_{\rm out}(r=r_0,t)&=& \frac{2 \alpha \rho_0 }{\pi r_0}\int_0^{\infty} d\gamma \: \frac{\gamma^2 e^{- \gamma^2 t}}{1+\gamma^2} \nonumber \\
& = &\frac{\alpha \rho_0}{r_0} \bigg[ \frac{1}{\sqrt{\pi \bar{t}}}-e^{\bar{t}} {\rm erfc}(\sqrt{\bar{t}})\bigg],
\end{eqnarray}
where ${\rm erfc}(x)$ is the complementary error function~\cite{abramowitz}. Using Eq.~(\ref{22}), we can expand the density of the outside at long times, and obtain the long-time behavior of the pressure which reads
\begin{equation}\label{23r}
\lim_{\bar{t}\to\infty} \bar{P}^{\rm out}_{3D}(\bar{t})=\frac{\bar{t}^{-\frac{3}{2}}}{2\sqrt{\pi}}.
\end{equation}
As for the 2D case, the time-dependent density inside and outside of the sphere can now be used to compute the force exerted on the sphere after the quench. This force is shown in Fig. \ref{fig4}(b) as a function of time. Figs.~\ref{fig6}(c) and \ref{fig6}(d) show that the force exerted on the spheres jumps to the value $\bar{F}=2$ very rapidly after the quench. As discussed in the previous section, this sudden jump is an interesting non-equilibrium effect associated with the curvature of the walls (more details can be found in the appendix \ref{ap2}). A comparison to Fig.~\ref{fignew} shows that the late time decay of the force towards its steady-state value follows $t^{-1}$ in 2D, but $t^{-3/2}$ in 3D. Accordingly, we can generalize these results for $d$-dimensional spherical geometries as $\bar F(\bar t) - {\bar F}(\bar{t}\to\infty)\propto \bar{t}^{-d/2}$ for large $\bar t$. 

\subsection{Quench of medium between two spheres}

As a final step, we consider a system of non-interacting Brownian particles confined between two spheres with radii $r_1$ and $r_2$, respectively; see Fig.~\ref{fig1} (d). The distance between the surfaces of the two spheres is defined as $\Delta r=r_2-r_1$. 
Using the solution found for the density in Eq.~\eqref{12}, subject to no-flux boundary conditions on the surface of two spheres, $\partial_r R(r)|_{r=r_1,r_2}=0$, we can find discrete values for the parameter $\beta^{\prime}$ from the following equation:
\begin{equation}\label{24}
\frac{\Delta r \beta_n^{\prime}}{\beta_n^{\prime^2} r_1 r_2+1} =\tan (\Delta r \beta_n^{\prime}),~~~n=1,2,3,\dots ~.
\end{equation}
The time-dependent solution for the density follows
\begin{eqnarray}\label{25}
&& \Delta\rho(r,t)= \sum_{n=1}^{\infty} c_n \frac{e^{-D \beta^{\prime^2}_n t} }{r} f_n(r)+c_0,\\
&&  f_n(r)=\sin(\beta^{\prime}_n r)+ a_n \cos(\beta^{\prime}_n r),
\end{eqnarray}
with 
\begin{equation}\label{26c}
a_n =
\frac{\beta^{\prime}_n r_1 -\tan(\beta^{\prime}_n r_1)}
{1+\beta^{\prime}_n r_1\tan(\beta^{\prime}_n r_1)}\equiv
\frac{\beta^{\prime}_n r_2 -\tan(\beta^{\prime}_n r_2)}
{1+\beta^{\prime}_n r_2\tan(\beta^{\prime}_n r_2)}.
\end{equation}

\begin{figure}[t!]
\centering
\includegraphics[width=0.99\linewidth]{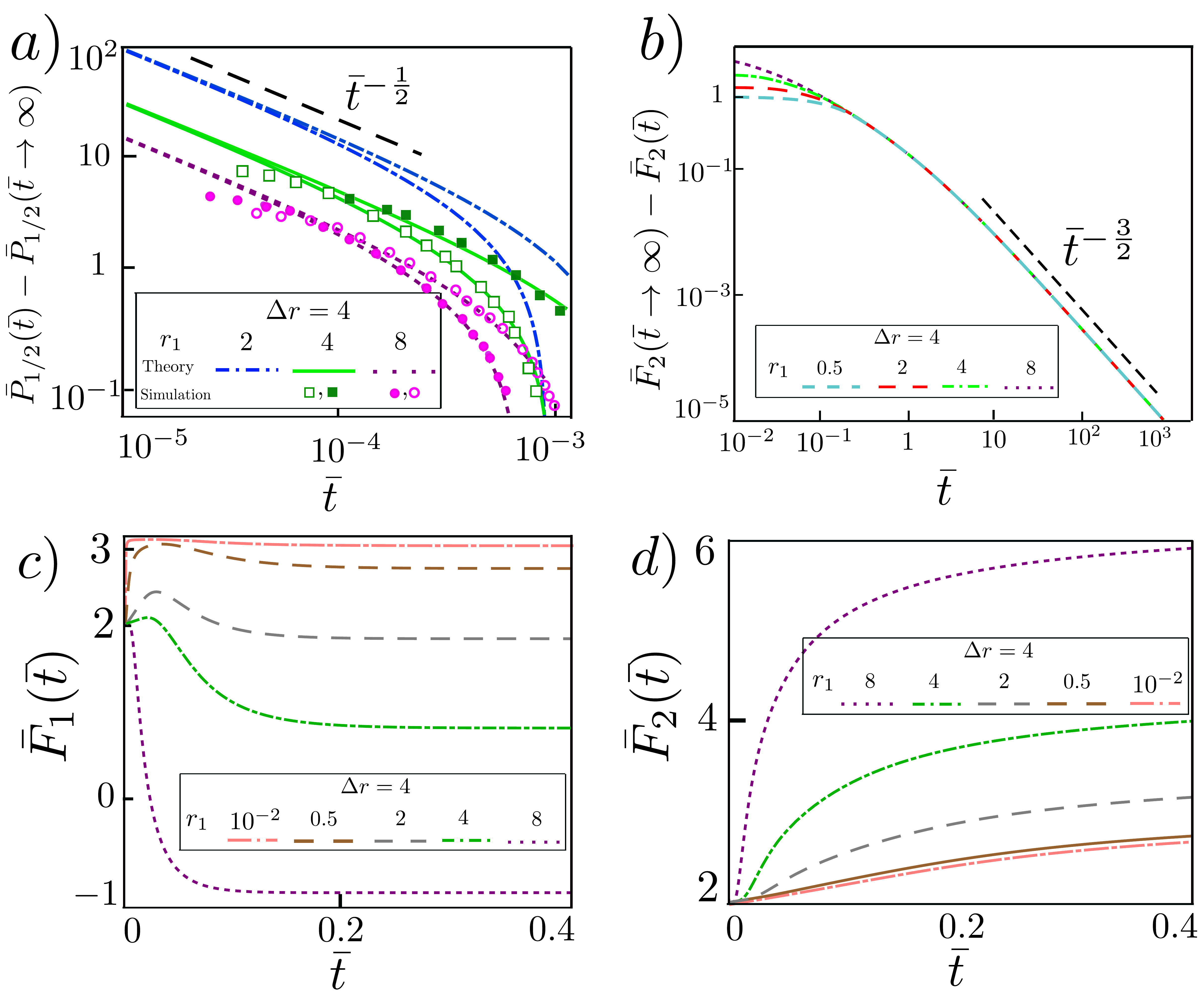}
\caption{
Time evolution of the pressure and forces exerted on the two spheres. (a) The solid lines and filled symbols (dashed lines and hollow symbols) correspond to the pressure $\bar P_1$ on the small sphere ($\bar P_2$ on the large sphere).
(b) Scaling behavior of the force on the large sphere upon approaching its steady-state value. Panels (c) and (d) show the forces exerted on the small and large spheres, respectively. Similar to Fig.~\ref{fig2cir}, forces are obtained from the analytical results.
}
\label{fig6}
\end{figure}
The initial condition for this problem has the form $\Delta\rho(r,t=0) = \alpha_3 \rho_0 \big[\delta(r-r_1)+ \delta(r-r_2)\big]$, where $c_0$ 
is fixed by calculating
\begin{eqnarray}\label{26}
&&\int_{r_1}^{r_2} r^2 \: dr \: \big[\frac{c_n}{r} f_n(r)+c_0 \big] \nonumber\\&&= \int_{r_1}^{r_2} r^2 \: dr \: \alpha \big[ \delta(r-r_1)+ \delta(r-r_2)\big].
\end{eqnarray}
One finds
\begin{equation}\label{26b}
c_0=\frac{3\alpha  \rho_0( r_1^2+ r_2^2)}{r_2^3-r_1^3}. 
\end{equation}
For the coefficients $c_n$, 
we need to compute
\begin{eqnarray}\label{27}
&& \int_{r_1}^{r_2} r \: dr \: \alpha \rho_0 \big[\delta(r-r_1)+ \delta(r-r_2)\big] f_m(r)  \nonumber \\&&=\int_{r_1}^{r_2}
r \: dr \:
\rho(r,{t}=0)f_m(r) ,
\end{eqnarray}
which gives
\begin{eqnarray}\label{28b}
c_m=2\alpha \rho_0 
\frac{r_2 f_m(r_2)+r_1 f_m(r_1)}
{r_2 f_m^2(r_2)-r_1 f_m^2(r_1)}.
\end{eqnarray}
To find the coefficients $c_m$, we have used the orthogonality condition
\begin{eqnarray}\label{orthosph}
\int_{r_1}^{r_2} dr \: f_m(r) f_n(r)= \frac{\delta_{m n}}{2} 
\big[ 
r_2 f_m^2(r_2)-r_1 f_m^2(r_1)
\big].
\end{eqnarray}
As is shown in Fig.~\ref{fig6}(a), the pressure exerted on the small and large sphere, $\bar{P}_1$ and $\bar{P}_2$ respectively, decay as $\bar{t}^{-{1}/{2}}$ at small times, for different values of $r_1$ and $r_2$.
The same scaling was observed for the 2D case  
and also for the single sphere as shown in Figs.~\ref{fig2cir}(a) and \ref{fig4}(a), respectively. 
Fig.~\ref{fig6}(b) shows that the force exerted on the large sphere approaches its steady-state as $\bar{t}^{-\frac{3}{2}}$ at long times.
Finally, Figs. \ref{fig6}(c) and (d) represent the time evolution of the forces acting on the two concentric spheres; these are qualitatively very similar to those of the two circles shown in Fig. \ref{fig2cir}. 
In particular, the forces suddenly jump to a finite value (${\bar F}_{1,2}=2$) and the force exerted on the small sphere experiences an overshoot for large enough values of $r_1$. The overshoot becomes more pronounced when $r_1$ is increased for a fixed $\Delta r$.

\section{Conclusions}\label{sec:conc}
We have studied systems of Brownian particles confined by surfaces in spherical geometries for $2$ and $3$ spatial dimensions, following a quench in temperature. For all cases considered, the analytical results were shown to be in quantitative agreement with our explicit Brownian dynamics simulations.

In particular, we calculated the time-evolution of the density analytically. This allowed us to find the dynamics of pressures and net forces acting on the boundaries. Short-time scaling of pressures on curved boundaries was shown to be insensitive to the dimension of the system, and universally decays as $t^{-1/2}$ dictated by effective a one-dimensional diffusion.
The observations made for forces, which are obtained as the difference of inside and outside pressures on a given boundary, are different. Unlike what is observed for flat boundaries~\cite{rohwer18}, the long-time scaling of the forces exerted on curved boundaries was shown to depend on the dimension of the system as $\bar{t}^{-{d}/{2}}$.
Furthermore, we showed that the curvature of the boundary differentiates diffusion of particles on the two sides of the boundary after the quench, i.e., particles close to the boundary and inside a sphere have less space to diffuse than particles close to the outside boundary. As a result, the post-quench force exerted on the curved boundaries, jumps to a constant value very rapidly; the constant depends on the dimension of the system. 

We have therefore demonstrated that the curvature of confining boundaries has an important (dimension-dependent) effect on non-equilibrium dynamics of an ideal fluid. Our results could, in principle, be used to compute forces acting on boundaries of droplets or bubbles in an ideal fluid. Additionally, the theoretical framework established here sets the basis for computing curvature corrections for forces on planar surfaces; this will be addressed in future work.

\acknowledgments
We thank P. Nowakowski for valuable discussions. Further, we thank S. Dietrich for funding a research visit of MRN at the MPI-IS in Stuttgart, during which this work was initiated. M. R. N. acknowledges the support of the Clarendon Fund Scholarship.
A.G.M. acknowledges financial support and the hospitality of the Leibniz Institute for Theoretical Solid State Physics (IFW Dresden) during his visit.

\appendix
\section{Boundary layer thickness $\alpha_d$}\label{ap1}

The number of particles before and after the quench is conserved; this constraint allows us to find the coefficient $\alpha_d$ as explained in Ref.~\cite{rohwer18}. For the circle we have
\begin{eqnarray}\label{s6}
\alpha_2 &&= \frac{1}{r_0}\int_{r_0}^{\infty} r \: dr \: [e^{-V_{\rm wall}/(k_B \: T_I)}-e^{-V_{\rm wall}/(k_B \: T_F)}]  \nonumber\\
&&\approx \sqrt{\frac{ \pi k_B}{2 \lambda}} (\sqrt{T_I}-\sqrt{T_F}).
\end{eqnarray}
We have used the approximation $\sqrt{\frac{2 k_B T_I}{\lambda}}, \sqrt{\frac{2 k_B T_F}{\lambda}} \ll r_0$ which means we have considered system sizes much larger than the characteristic width of the boundary layer.
To calculate the coefficient $\alpha_3$, we need to integrate the initial and final density in $3D$. We use the same approximation for the system size and find:
\begin{eqnarray}\label{s7}
\alpha_3 &&= \frac{1}{r_0^2}\int_{r_0}^{\infty} (e^{-V_{\rm wall}/(k_B \: T_I)}-e^{-V_{\rm wall}/(k_B \: T_F)})  r^2 dr   \nonumber\\
&&\approx \sqrt{\frac{ \pi k_B}{2 \lambda}} (\sqrt{T_I}-\sqrt{T_F}),
\end{eqnarray}
which is equal to the coefficient $\alpha_2$. For this reason we have used $\alpha$ instead of $\alpha_d$ in the main text.

\section{Explicit derivation of the jump in the force}\label{ap2}
The (dimensionless) force acting on the curved boundaries after the quench jumps very rapidly to a constant value that depends on the dimension of the system. The constant value is equal to one and two for a circle and a sphere, respectively. Here, we show that the force exerted on a sphere is indeed equal to zero at $\bar{t}=0$, but very rapidly reaches a value $\bar{F}_{3 D}=2$, whereafter it evolves towards its steady state value. We also discuss the reason for this jump, which is absent in a system with flat boundaries. 
We consider the initial condition for the excess density as
  \begin{equation}\label{s1}
  \Delta\rho(r,\bar{t}=0) =
    \begin{cases}
       \frac{\alpha \rho_0 }{ \epsilon r_0}, & |\frac{r}{r_0}-1|<\frac{\epsilon}{2} \\
      0, & \text{otherwise}.
    \end{cases}
  \end{equation}
We take $\epsilon$ to be very small (so that the expression approximates a $\delta$ function), and use this initial condition to find the unknown coefficients in the density expansion in Eqs.~\eqref{15} and \eqref{20}. The density on the interior of the sphere at $t=0$ is then
\begin{eqnarray}\label{s2}
&&\Delta \rho_{\rm in}= \frac{\alpha\rho_0}{ r_0} \bigg[
3+\frac{2}{\epsilon}\sum_{m=1}^{\infty}\frac{\Lambda_m-\beta_m \epsilon \cos(\beta_m \epsilon)}{\beta_m^3 }
\bigg]
,\nonumber \\
&&\Lambda_m=\sin (\beta_m \epsilon)[1+\beta_m^2 (1-\epsilon)],
\end{eqnarray}
where the coefficients $\beta_m$ are the positive solutions of Eq.~\eqref{14}. The summation in Eq.~\eqref{s2} can be approximated with an integral based on the Euler–Maclaurin formula and the fact that $\epsilon\ll 1$ \cite{apostol99}. To this end, we first add 
$\epsilon-\epsilon^2+\epsilon^3/3$, 
which can be thought of as the $m=0$ term of the summation (corresponding to $\beta_0\to 0$). Then, noticing that the distance between successive roots $\beta_{m}$ of Eq.~\eqref{14} is very close to $\pi$ 
for large $m$, we use the conversion $\sum_{m=0}^{\infty}\to\int d\beta/\pi$ which yields 
\begin{eqnarray}\label{s4}
&&\Delta \rho_{\rm in}= \frac{ \alpha \rho_0}{r_0} \bigg[ 3+\frac{2 }{\epsilon }\int_0^{\infty}\frac{d\beta}{\pi}\: \frac{\omega(\beta)-\beta \epsilon \cos(\beta \epsilon)}{\beta^3 }\nonumber \\
&&\qquad\qquad -
\frac{2}{\epsilon}(\epsilon-\epsilon^2+\frac{\epsilon^3}{3})\bigg]+{\cal O}(\epsilon),\nonumber \\
&&\omega(\beta)=\sin (\beta \epsilon)[1+\beta^2 (1-\epsilon)],
\end{eqnarray}
and eventually 
\begin{eqnarray}\label{s4p}
&&\Delta \rho_{\rm in}= \frac{ \alpha \rho_0}{r_0} \frac{1}{\epsilon}
+{\cal O}(\epsilon).
\end{eqnarray}
Similarly and from Eq.~\eqref{20}, the density exterior to the sphere at $t=0$ is found to be
\begin{eqnarray}\label{s3}
&&\Delta \rho_{\rm out}= \frac{\alpha \rho_0}{r_0} 
\frac{2}{\epsilon} \int_0^{\infty}\frac{d\beta}{\pi}\:\frac{\Omega(\beta)-\beta \epsilon \cos\beta \epsilon}{\beta\big(1+\beta^2\big)}
\bigg],\nonumber \\
&&\Omega(\beta)=\sin \beta \epsilon[1+\beta^2 (1+\epsilon)],
\end{eqnarray}
which, after evaluating the integral, leads to 
\begin{eqnarray}
\Delta \rho_{\rm out}= \frac{\alpha \rho_0}{r_0} \frac{1}{\epsilon}.
\end{eqnarray}
Therefore the force exerted on the sphere immediately following the quench is proportional to $\Delta\rho_{\rm in}-\Delta \rho_{\rm out}\propto \epsilon$ which means the force is equal to zero at $\bar{t}=0$ for $\epsilon\to0$. 

\par
As time evolves from zero, the force increases very rapidly to the value $\bar{F}=2$, after which the approach to steady state continues as discussed in the main text. Assuming $(\epsilon/r_0)^2 \ll \bar{t}$, we first take the limit of $\epsilon \to 0$ and then the limit of $\bar{t} \to 0$ in Eqs.~\eqref{19} and \eqref{20} from which we obtain
\begin{eqnarray}\label{s5}
&&\Delta \rho_{\rm in}(r=r_0,\bar{t}\to0)=\frac{\alpha\rho_0}{r_0}(3+\sum_{m=1}^{\infty} 2),\\
&&\Delta \rho_{\rm out}(r=r_0,\bar{t}\to0)= \frac{2\alpha\rho_0}{r_0}\int_0^{\infty}
\frac{d\beta}{\pi}\:
\frac{\beta^2 }{1+\beta^2 }.\quad
\end{eqnarray}
Converting the summation in internal density to an integral in the same way explained above, we can then calculate the force exerted at the sphere for finite and small values of the time as:
\begin{eqnarray}\label{ss5}
\bar{F} &=& \frac{r_0}{\alpha \rho_0}\big[\Delta \rho_{\rm in}(r=r_0,\bar{t}\to0)-\Delta \rho_{\rm out}(r=r_0,\bar{t}\to0)\big]\nonumber\\
&=& 1+\frac{2}{\pi}\big(\int_0^{\infty} d\beta-\int_0^{\infty}d\beta\:\frac{\beta^2 }{1+\beta^2 }\big)=2,
\end{eqnarray}
which matches very well with the inset of Fig.~\ref{fig4}(b). This rapid change in the force acting on the curved surface of the sphere is absent for flat boundaries. In fact, due to the curvature of the boundary, particles on the inner surface of the sphere are more confined comparing to their external counterparts. As a result, at a finite time, the exterior particles diffuse and spread in space more rapidly. This causes a rapid increase in the force acting on the boundary. A similar process occurs for the force acting on the circle in 2D.

\bibliography{nkrm.bib}

\end{document}